\documentclass[pra,twocolumn,preprintnumbers,amsmath,amssymb]{revtex4}

\usepackage{graphicx}
\usepackage{dcolumn}
\usepackage{bm}
\usepackage{color}
\usepackage{epstopdf}
\usepackage{float}
\usepackage{ulem}

\renewcommand{\em}{\textit}

\begin{document}



\title{Analogue black hole lightring and its resonances in an optical fiber}
\author{R. Terrier$^{1}$, J. Fatome$^{1}$, B. Kibler$^{1}$, and T. Torres$^{1,*}$}
\affiliation{$^{1}$ Universit\'e Bourgogne Europe, CNRS, Laboratoire Interdisciplinaire Carnot de Bourgogne ICB UMR 6303, 21000 Dijon, France}
\affiliation{$^{*}$ Corresponding author: theo.torres@ube.fr}

\date{\today}

\begin{abstract}
We demonstrate the complete observation of a black-hole lightring resonance in an optical analogue. 
To do so, we generalize the concept of the photon sphere to non-stationnary systems by showing that that the lightring is associated to a spatiotemporal caustic and that its complex frequency is determined by the propagation and divergence of this caustic.
For an optical soliton acting as an analogue black-hole potential, we predict a characteristic spectral plateau whose boundaries are set by the real part of the lightring frequency, while its spectral decay is governed by the imaginary part. 
Using a pump-probe experiment in a 5-km-long dispersion-shifted fiber, we observe these signatures for two distinct pump wavelengths. 
The measured spectra agree with numerical simulations and with the predicted lightring-mode profile, providing direct access to both the oscillatory and decay properties of the resonance. 
Our results establish a direct experimental connection between black-hole lightrings, spacetime caustics, and quasinormal-mode spectroscopy in a dispersive analogue system.
\end{abstract}


\maketitle

{-- \it Introduction.} 
Resonances are fundamental in all fields of Science. They encapsulate the intrinsic properties of a system through their spectral signatures and, as a consequence, provide a powerful means of extracting information. This principle forms the basis of spectroscopy, where the spectrum of resonances can be used to characterize a system through the observation of its response to a perturbation. Spectroscopic techniques have found applications across a broad range of fields, including optics, acoustics, and seismology~\cite{Brady,Dorfman,Veksler,Bath}. More recently they also became an important tool to study astrophysical black holes for which the radiation emitted following a perturbation can provide information about the properties of this elusive system that cannot be accessed directly~\cite{Cabero,Berti2026}.

Black-hole spectroscopy is closely connected to the study of quasinormal modes (QNMs), which describe the characteristic damped oscillations produced by a perturbed black hole~\cite{Nollert,Kokkotas1999,Berti2009,Konoplya2011}. 
These modes are determined by global properties of the spacetime and, in particular, by the boundary conditions. 
As a consequence, QNMs encode information about the geometry of the black hole from its horizon up to infinity. 
It is precisely the global nature of QNMs that makes their theoretical analysis and experimental observation challenging. 
In particular, the spectral properties of QNMs can exhibit sensitivity to small perturbations of the geometry~\cite{Jaramillo2021,Destounis2024}, while experimentally accessing a mode whose definition involves the complete evolution of the field remains a difficult task.

A way to gain insight into the above problems is, as often, to use approximations. In the context of black hole QNMs physics, this is provided by the concept of the black-hole lightring. In stationary black-hole spacetimes, the lightring is associated with an unstable circular orbit of null geodesics and can be identified from an unstable equilibrium of the radial potential~\cite{Anderson2002}. In the semiclassical regime, the properties of this orbit are directly related to those of the corresponding QNMs, providing a local description of resonances that are otherwise defined by global boundary conditions~\cite{Ferrari1984,Schutz1985,Cardoso2009}. This connection offers an attractive route toward studying black-hole resonances through the dynamics of waves near the lightring. The price to pay for this approximation is to go from a set of modes which encapsulate the global properties of the system to modes which depend solely on local properties. However, one can argue that this is a relatively low price to pay considering that the concept of lightring modes can be extended to dispersive systems and still be used to perform spectroscopy~\cite{Torres2018waves,Torres2019spectro}.

The lightring concept is particularly interesting in analogue systems, where the dynamics of waves can reproduce mathematical structures associated with curved spacetimes~\cite{unruh,visser,AGreview,Jacquet}. Analogue systems provide a versatile platform for this purpose, not only can they mimic fundamental gravitational physics~\cite{Philbin,Weinfurtner,Euve,Torres2017,Steinhauer2019,svanvcara2024}, they also enrich these phenomena with specific condensed matter effect such as dispersion~\cite{Jacobson,Unruh1995} or additional degrees of freedom such as vorticity~\cite{Patrick2018}.

In this Letter, we propose to investigate the concept of lightring mode in an optical analogue system both theoretically and experimentally. Our results are twofold: i) We relate lightrings to spatio-temporal caustics, offering a new perspective and set of tools on this fundamental object and ii) we demonstrate experimentally the presence of lightring modes in an optical fiber.

\medskip
\noindent
{-- \it From black hole lightring to spacetime caustics.}
While condensed matter and black holes share an astonishing amount of properties, most of them based on concrete mathematical foundations, it is important to recall that they also differ in many ways. 
Most relevant in the context of optics, is the fact that massless fields around black hole propagate without dispersion. 
This is certainly not the case in optics, and therefore reformulating gravity inspired concept in a language that is applicable to dispersive systems is warranted.

\begin{figure}[h!]
    \centering
    \includegraphics[scale = 0.6]{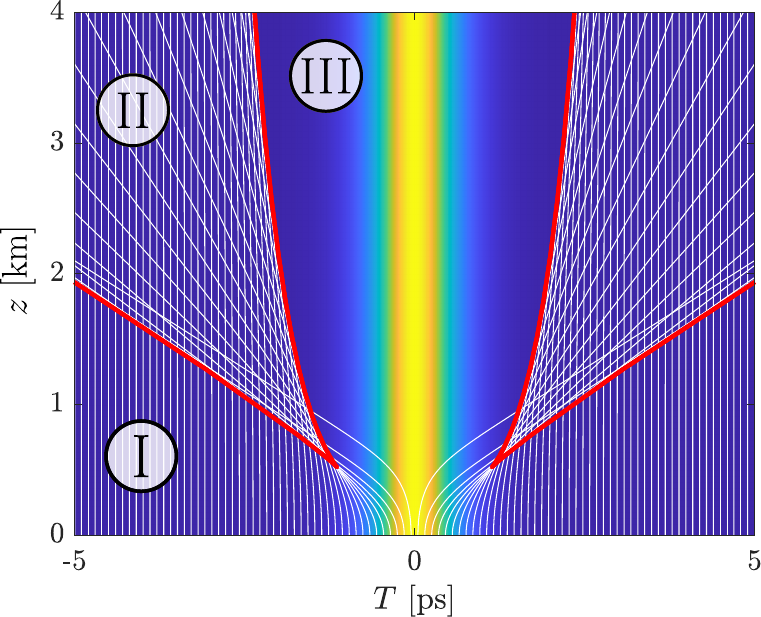}
    \caption{\textbf{Spatiotemporal caustic from the analogue lightring.} The background shows the $z$-independent optical potential created from a fundamental soliton. The white curve depicts the ray trajectories associated with a initially GVM probe field. The rays are deflected which result in the spatiotemporal caustic illustrated by the red curve. The caustic are composed of an inner branch and an outer branch moving away at a speed defined entirely from the properties of the soliton. Region I, II and III corresponds to regions which can be reached by either 1, 2 or no rays.}
    \label{fig:caustic}
\end{figure}

As stated above, one important feature of stationary black holes is the existence of the photon sphere. 
This object is defined through an inspection of the \textit{radial} behaviour of the geodesic in a \textit{stationnary} black hole spacetime and can be generalized to dispersive systems~\cite{Torres2018waves}.
Below, we propose to generalize further the concept of the photon sphere and to establish a link between the lightring of (analogue) black holes, and a \textit{spacetime} caustic in a dispersive medium.
Intuitively, this link is made from the fact that a one dimensional barrier of a stationary system, act as a spatiotemporal diverging lens.
We explicitly illustrate this link by considering the case of a weak optical probe pulse scattering of an optical analogue black hole lightring.
The evolution equation for a group velocity matched (GVM) weak pulse envelop, $a(T,z)$ can be cast in the form of the linear Schrodinger equation in a potential~\cite{agrawal,Burgess2024}
\begin{equation}\label{eq:WE}
    -i\frac{\partial a}{\partial z} + \frac{\beta_2^{\rm GVM}}{2}\frac{\partial^2a}{\partial T^2} = V(T,z) a(T,z).
\end{equation}
In the above equation, we have neglected for simplicity the higher order dispersion coefficients, and assume that the optical potential is that of a solitonic pump pulse, i.e. $V(T,z) = 2\gamma P_s{\rm sech}^2\left( T/T_S\right)$, with $\gamma$ the nonlinear coefficient, $P_S$ the peak power of the soliton, $T_S$ is temporal width and $\beta_2^{GVM}$ is the group velocity dispersion at the GVM wavelength. We work in a coordinate system $(z,T)$ which is comoving with the soliton pulse such that, in this frame, the pump profile is independent of $z$.
The weak probe is assumed to be travelling at the same group velocity than the soliton but in the normal dispersion regime, i.e. $\beta_2^{\rm GVM} >0$.
We stress here that the $z$ and $T$ coordinates are the analogue of the time and the ``tortoise" coordinates in a black hole spacetime. In those coordinate, the analogue black hole horizon is located at $T\rightarrow -\infty$.
From the wave equation \eqref{eq:WE}, we can construct a Hamiltonian, $\mathcal{H}$, governing the propagation of rays in the potential $V$, the analogue of the geodesics in a curved spacetime~\cite{Torres2018waves,supplement,Torres2026OL}. In the case of a $z$-independent Hamiltonian, one can define the lightring from the conditions $\partial_T\mathcal{H} = 0$ and $\partial_\Omega \mathcal{H} = 0$. Using the fact that the lightring is an unstable equilibrium point, we can further define its associated lightring wavenumber $k_z^\star \in \mathbb{C}$~\cite{supplement}. The real part of $k_z^\star$ is associated to the wavelength of a wave located on the lightring and its imaginary part to its exponential decay as the wave leaks out of it. In the case of a pump field corresponding to a fundamental soliton, the lightring wavenumber is given by
\begin{equation}
    k_z^\star = 2\gamma P_S - i \Lambda/2,
\end{equation}
where $\Lambda = \sqrt{4\gamma\beta_2^{\rm GVM}P_S/T_S^2}$. 
\\
We now highlight that these quantities have an equivalent interpretation in terms of a spacetime caustic.
Fig.~\ref{fig:caustic} shows the trajectories of the rays launched at the comoving frequency with the soliton.
The spacetime can be decomposed into three regions, caracterized by the number of rays reaching each regions and separated by a cusp caustic.
In region I, only a single ray reaches every spatial point. This is due to the finite deflection of the rays. The spectral content of region I will be mainly that of the initial continuous probe field.
Region III corresponds to the spatial region which cannot be reached by any rays due to the diverging nature of the soliton. There is therefore no spectral power contained within this region.
In region II, two rays can reach any spatial points: i) an initial ray originally close to the maximum of the soliton which will be strongly deflected and ii) a ray initially further away from the soliton which is weakly deflected. Those two rays will interact leading to interference in this region. The spectral content of region II is more intricate but in its bulk, it contains the cross phase modulation of the pump onto the probe field.
Those three regions are separated by cusp caustics, each composed of two branches.

From the ray analysis, we see that the two branches of the caustic emerge from different part of the potential. In particular the inner caustic arises from the ``tail" of the soliton while the outer one can be traced back to the maximum of the potential, i.e. the lightring. 
It is the latter that we shall focus on.
The outerbranch moves away from the soliton with a certain inverse group velocity $\Delta \beta_1^\star$, which allows for a clear interpretation of the oscillation of the lightring modes: the lightring modes propagate at the frequency which travel at the speed of the caustic outerbranch.
The imaginary part of the lightring frequency is associated with the spatial location of the caustic, i.e. how quickly the ray diverge away from the potential.
In the model considered here, the potential is governed by two parameters: the peak power and width of the soliton. 
Since we are considering the propagation in the $(z,T)$ plane, and the initial position of the rays are parameterized by a single state parameters, employing the classification of caustic~\cite{Berry1980}, the lightrings are associated with the elementary caustic of codimension 4: the butterfly caustic~\cite{Torres2022wave}. 
Since every caustic belonging to the same elementary class can be smoothly deformed into one another, one can recover the properties of the lightring modes from the caustic shape.

\begin{figure}[!h]
    \centering
    \includegraphics[scale = 0.65]{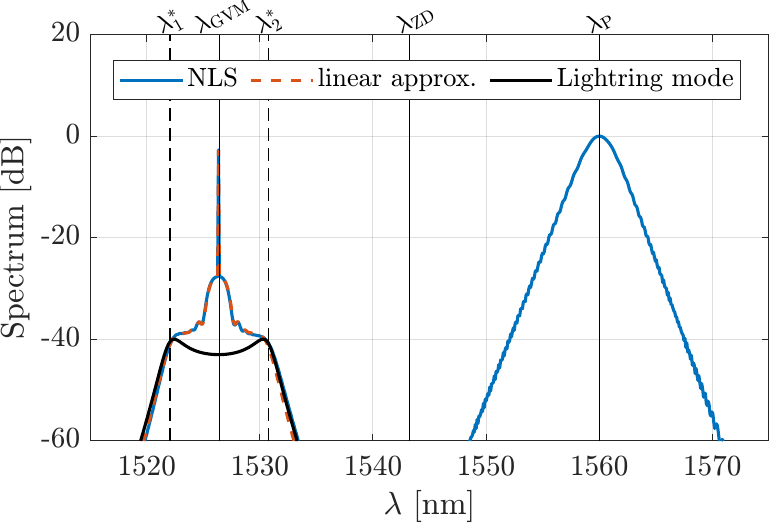}
    \caption{\textbf{Spectral signature of the lightring resonance.} Numerically simulated spectrum from the linear wave equation Eq.~\eqref{eq:linear_WE} and the GNLSE compared with the lightring mode spectrum. The solid blue (dashed red) curve show the spectrum at the output of a $5$-km-long dispersion shifted fiber whose parameters correspond to the experimental configuration detailed in the main text. The spectrum exhibit a clear plateau covering a spectral range $(\lambda_1^\star,\lambda_2^\star)$ given by the lightring properties. Furthermore, the spectral decay matches extremely well with the spectrum of the lightring mode defined in Eq.~\eqref{eq:LR_mode}}
    \label{fig:full_simu}
\end{figure}

\medskip
\noindent
{-- \it Lightring mode spectrum.} As we have previously discussed, the lightring is characterised by its evolution in $z$ which is governed by the complex wavenumber $k_z^\star$. This signal, similar to the ringdown response emitted by a perturbed black hole, is experimentally difficult to access in a optical fiber experiment. However in such platforms we can access with high precision the temporal spectrum of light at the output of the fiber. 
To do define the temporal spectrum, or equivalently the temporal profile of the lightring resonance, we first relate the lightring wavenumber to a complex temporal frequency $\Omega^\star$ via the dispersion relation 
\begin{equation} \label{eq:main_freq}
    \Omega^\star = \sqrt{\frac{2 k_z^\star}{\beta_2^{\rm GVM}}}.
\end{equation}
The generalisation to higher dispersion is done in~\cite{supplement}.
We make use of the correspondence between the lightring mode and the fundamental quasinormal modes~\cite{Cardoso2009} and anticipate that it holds not only for the complex frequencies but also for their respective mode profile. 
The temporal profile of the fundamental QNM is given by \mbox{$
a_{\rm QNM}^{(n=0)}(T) \propto {\rm cosh}\left( T/T_S\right)^{-i\Omega_{\rm QNM}T_S}$}~\cite{Burgess2024,Besson2025}. 
From this result, we define the temporal profile of the lightring mode as
\begin{equation}\label{eq:LR_mode}
    a_{\rm LR}(T) \propto {\rm cosh}\left( \frac{T}{T_S}\right)^{-i\Omega^\star T_S}.
\end{equation}
The above mode has the form of a right-moving wave as $T\rightarrow +\infty$ and of a left moving wave as $T\rightarrow -\infty$.
To look for the presence of lightring mode in our optical analogue, we numerically evolve a global perturbation, inspired by~\cite{Torres2020}, composed of a continuous probe field at the group velocity matched frequency, using both the linear evolution and the full generalised nonlinear Schr\"odinger equation (GNLSE)~\cite{agrawal}.
Fig.~\ref{fig:full_simu} shows the spectrum at the output of a $5$-km-long fiber corresponding to the experimental condition described below. 
Firstly, we see that the linear evolution of the probe field (dashed red curve) agrees extremely well with the GNLSE (solid blue curve), justifying the framework of linear perturbation theory. 
We see that the probe field spectrum reveal a distinct spectral plateau in log scale around the GVM wavelength $\lambda_{\rm GVM}$. 
The plateau ranges from $\lambda_1^\star$ to $\lambda_2^\star$ which corresponds to the lightring wavelength defined via $\lambda_{1,2}^\star = 2\pi c/(\omega_{\rm GVM} \pm \Omega^\star)$. 
We further see that the decay of the plateau beyond the $(\lambda_1^\star,\lambda_2^\star)$ range agrees remarkably with the temporal spectrum of the lightring modes Eq.~\eqref{eq:LR_mode}.

\medskip
\noindent
{-- \it Observation of resonances.}
Equipped with the above tools, we now perform a pump/probe experiment analogous to the interaction between a scalar field in the exterior of a black hole geometry.
The setup is composed of a mode-locked fiber laser delivering pulses with nearly ${\rm sech}^2$ shape at a 40 MHz repetition rate and a temporal full width at half maximum (FWHM) around 0.7 ps with an adjustable carrier wavelength between $1540\,{\rm nm}$ and $1560\,{\rm nm}$. 
In the following, we perform our experiment at two pump wavelengths: i) $1555\,{\rm nm}$ and ii) $1560\,{\rm nm}$.
The pump field is combined with a continuous wave probe before being injected into a $5$-km-long dispersion shifted fiber (DSF) with a zero dispersion wavelength at $1543\,{\rm nm}$. 
The nonlinear coefficient $\gamma$ is taken to be $1.8\,{\rm{W}}^{-1}.\rm{km}^{-1}$ and the dispersion coefficients are $\beta_2^P \approx -1.52 \ \rm{ps^2.km^{-1}}$, $\beta_3^P \approx 0.12\ \rm{ps^3.km^{-1}} $ and $\beta_4^P \approx -0.0006\ \rm{ps^4.km^{-1}}$ at the pump wavelength \mbox{$\lambda_P = 1560\ {\rm nm}$}~\cite{endmatter}.
The wavelength of the probe wave is set close to the group velocity matched wavelength of $1531.5\,{\rm nm} $ ($1526.5\,{\rm nm} $) when the pump is centred at $1555\,{\rm nm}$ ($1560\,{\rm nm}$).
The pump parameters are carefully chosen to ensure simultaneously that: i) the spectral width of the pump is not too large in order to avoid the effect of Raman scattering and such that it remains in the anomalous dispersion regime and ii) the pulse is sufficiently short to stimulate the emission of lightring resonances. 

\begin{figure}[!h]
    \centering
    \includegraphics[scale = 0.7]{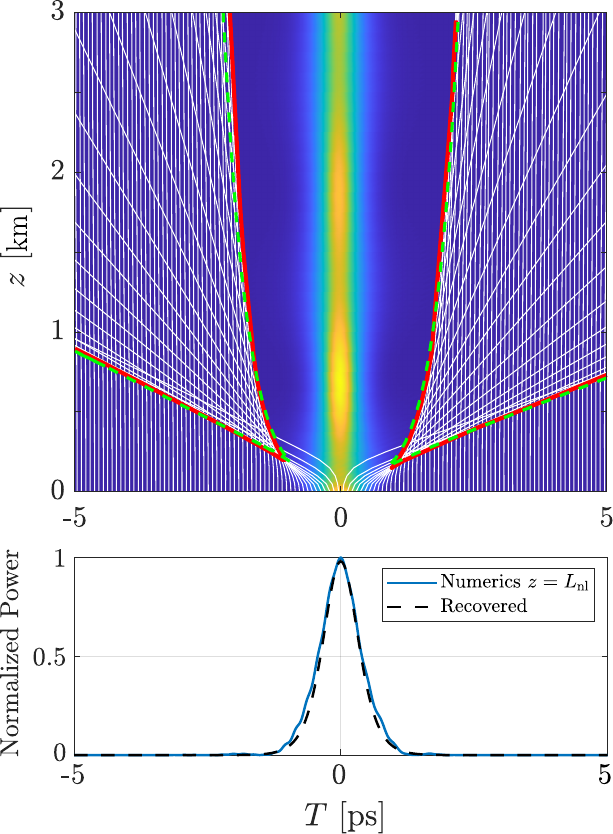}
    \caption{\textbf{Ray tracing on experimental pump pulse (top) and recovered pump pulse (bottom).} \underline{Top panel:} The background color represents the normalised intensity of the pump field. The white lines depicts the ray trajectory of the wave initially at the GVM frequency $\Omega=0$. The rays are deflected by the optical potential and form a caustic illustrated by the red curve. The green dashed curve corresponds to the caustic generated by a fundamental soliton that best matches the caustic from the spatiotemporal pump field. \underline{Bottom panel:} Comparison between the numerical pump pulse evaluated at $z=L_{nl}$ and the fundamental soliton associated with the best matched caustic.} 
    \label{fig:ray_exp}
\end{figure}

We characterise the main laser output and numerically simulate its propagation through the various optical components of the set-up (see~\cite{endmatter}) which serve as the initial condition for the propagation in the DSF.
During this short propagation, the pump field acquires a temporal chirp which has to be compensated by an increased input power as compared to the one of a fundamental soliton. 
For example, when the pump wavelength is centred around $1560\ {\rm nm}$, the measured FWHM is about $670\ {\rm fs}$ and the input power is $P_S \approx 6.7\ {\rm W}$ which is slightly above the input power of the fundamental soliton. The evolution of the pump field intensity in the DSF is shown in Fig.~\ref{fig:ray_exp}. We can see that peak power oscillates slowly during the propagation but that the overall temporal shape remains similar across the entire fiber.
These oscillations implies that the optical potential is $z$-dependent and therefore one cannot apply the standard definition of lightring. However, using the eikonal approach, it is still possible to define and characterise a spacetime caustic. This is illustrated by the white rays in Fig.~\ref{fig:ray_exp} which forms the caustic depicted by the red curves. Making use of the structural stability of caustic, we can look for the $z$-independent ${\rm sech}^2$ optical potential that produces a caustic that best match the one generated by our evolving pump field. 
This caustic is obtained for $P_S \approx 5.2\ {\rm W}$ and a FWHM of about $780 \ {\rm fs}$. 
Those parameters correspond to a pulse that matches well the pump field after having propagated for a distance corresponding to the nonlinear length $L_{nl} = (\gamma P_S)^{-1}$. 
The same observation can be made in the case $\lambda_P = 1555 \ {\rm nm}$. 

\begin{figure}[!h]
    \centering
    \includegraphics[scale = 0.7]{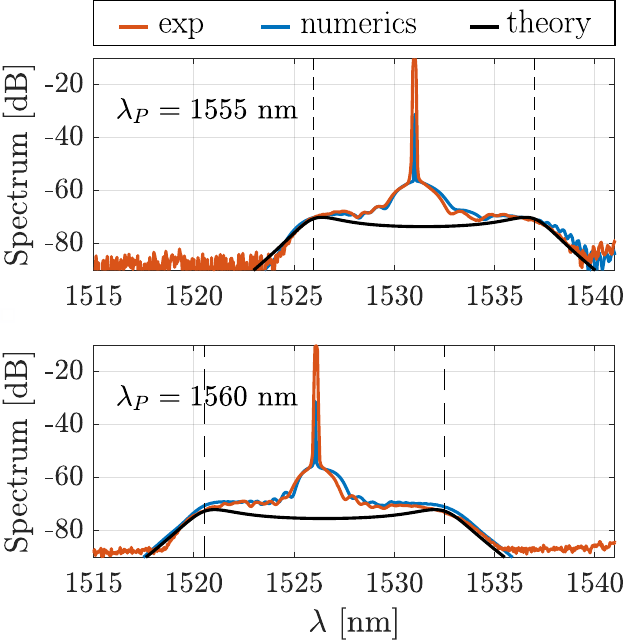}
    \caption{\textbf{Experimental signature of analogue lightring resonances.} Comparison between the measured (orange), simulated (blue) and theoretical (black) spectrum. The measured and simulated spectra are in excellent agreement and exhibit a spectral plateau spanning a frequency range defined by the lightring frequencies (equivalently wavelength) represented by the vertical dashed line. The spectral decay beyond the plateau is in good agreement with the theoretical spectrum of the lightring mode given by Eq.~\eqref{eq:LR_mode}. The top and bottom pannels correspond to different pump central wavelength, namely $\lambda_P = 1555 \ {\rm nm}$ and $\lambda_P = 1560 \ {\rm nm}$ respectively.}
    \label{fig:simu_vs_exp}
\end{figure}

After having identified the caustics from which we can construct the lightring temporal frequencies and modes profiles according to Eqs.~(\ref{eq:main_freq},\ref{eq:LR_mode}), we now turn our attention to the experimental spectral signature of the lightring resonances in our optical platform. 
We inject a continuous GVM probe field together with the pump and record the spectrum at the output of the fiber. 
Figure.~\ref{fig:simu_vs_exp} shows the measured spectrum (orange curve) together with the numerical simulation (blue curve) and the lightring mode spectrum (black curve). 
We can see that the probe spectrum exhibit the characteristic shape described previously, i.e. a broad peak centred at the GVM wavelength surrounded by a spectral plateau that spans between the lightring wavelength $(\lambda_1^\star,\lambda_2^\star)$. 
The spectral decay beyond the plateau is in excellent agreement with the spectrum of the lightring mode constructed from the caustic shape. 
We stress here that the agreement relies on both the real and imaginary part of the lightring frequency. Previously, only the real part of such modes was measured~\cite{Torres2020}, and therefore our measurement is the first complete observation of an analogue lightring resonance.

\medskip
\noindent
{-- \it Conclusion and perspectives.} 
In this work, we have investigated the resonant properties of an optical analogue of a black-hole lightring.
On the theoretical side, we have enriched the concept of black hole lightring which appears naturally in stationnary systems by showing that this unstable equilibrium point gives rise to a space-time caustic. 
This semiclassical picture provides a direct interpretation of the lightring oscillations as the propagation speed of the caustic. One of the main advantage of this picture is that it allows for the generalisation of the definition of the lightring frequency to non-stationnary systems. Using this correspondence, we constructed the temporal profile of the lightring mode and obtained its expected spectral signature. Numerical simulations based on both the linearized propagation equation and the GNLSE showed that the probe spectrum develops a characteristic plateau whose boundaries are determined by the lightring frequencies.
\\
On the experimental side, we have shown that analogue black hole resonances are readily accessible in rather minimalist set-ups built around commercially available sources. The measured spectra are in good agreement with both numerical simulations and the theoretical lightring-mode spectrum, providing an experimental observation of the analogue lightring resonance through both its oscillatory and decay properties. 
\\
Our results open several promising avenues of research, from investigating and applying more tools from semiclassical theory to the description of lightring resonances~\cite{Heller,Jaramillo2022}, to studying of analogue black hole resonances in non-stationnary systems, while paving the way for the realisation of the analogue black hole spectroscopy method.

\medskip
\noindent
{-- \it Acknowledgements.} TT. would like to wholeheartedly thank J.-L. Jaramillo and J. Besson for decisive discussions on the construction of the lightring mode profile, C. Finot for inspiring conversations on optical shock waves as well as P. Béjot, J. Carbillet, F. K\"onig, E. Lucas and S. Patrick for insightful exchanges.


\newpage
\baselineskip 11pt

\onecolumngrid
\section*{\Large End Matter}
\section*{\large Experimental set-up}

\begin{figure}[!h]
    \centering
    \includegraphics{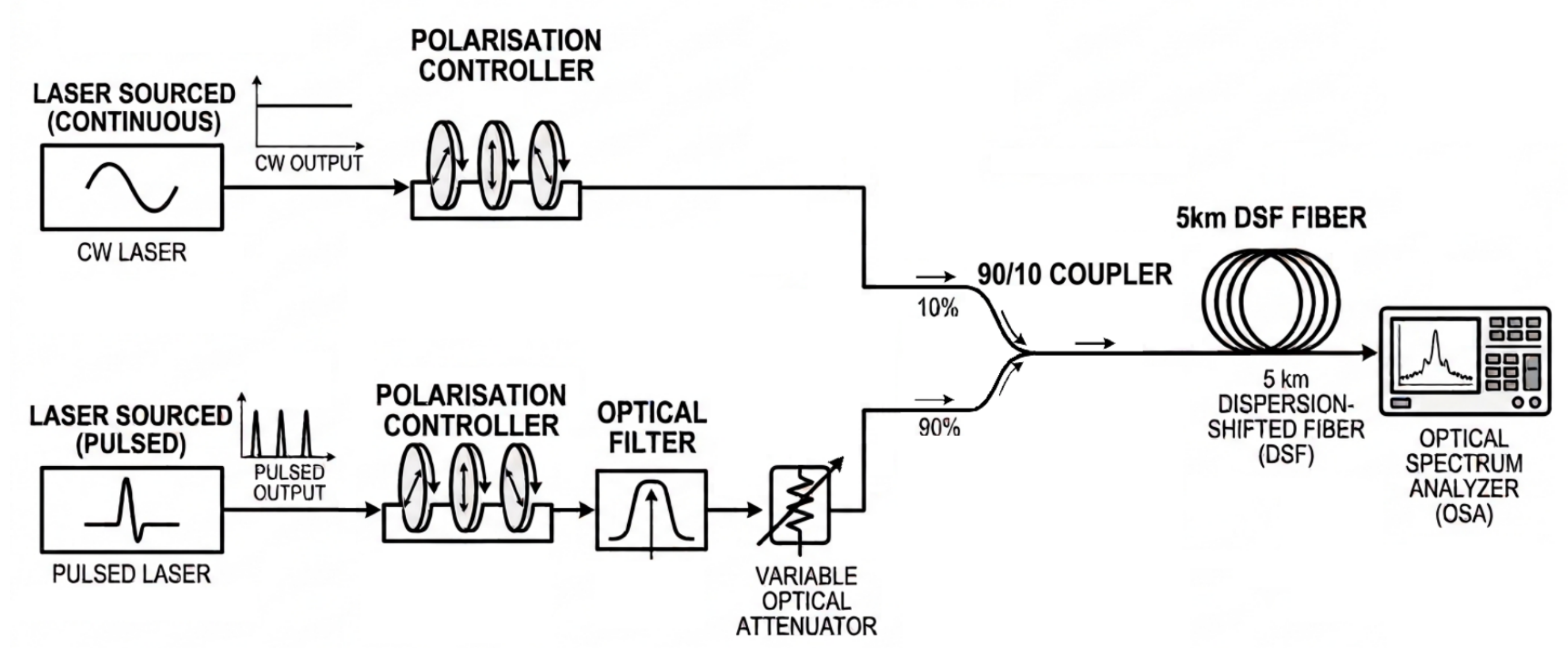}
    \caption{\textbf{Experimental set up.} A continuous probe field is combined with sub-ps pulses delivered by a mode-locked fiber laser before being injected in a $5$-km-long dispersion shifted fiber (DSF). An tunable optical filter and a variable optical attenuator are used at the output of the fiber laser in order to shape the pump field to ressemble a fundamental soliton. The output spectra are recorded at the output of the fiber with an optical spectrum analyzer.}
    \label{fig:set_up}
\end{figure}

\begin{figure}[!h]
    \centering
    \includegraphics[width = 0.7\textwidth]{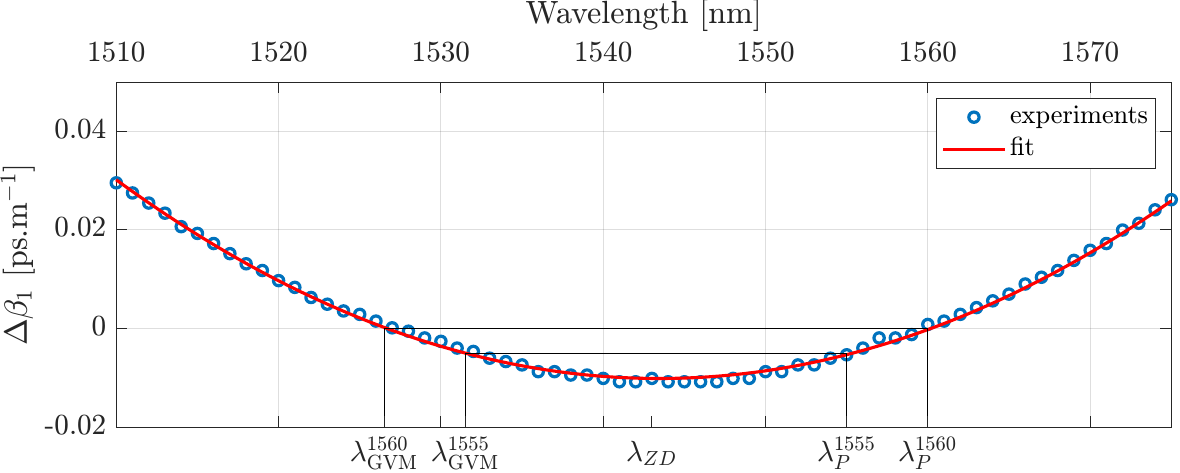}
    \caption{\textbf{Experimental measurement of the group velocity dispersion.} The blue circle represent the measurement group velocity dispersion via a time-of-flight measurement. The red curve is a third order polynomial fit which we allow us to extract the second, third and forth order dispersion coefficients given in the main text.}
    \label{fig:Db1}
\end{figure}

We detail below the experimental set up used to observance the optical lightring resonances discussed in the main text.
Fig.~\ref{fig:set_up} shows the various element of the fiber based platform. It is composed of a mode-locked fiber laser delivering pulse with nearly ${\rm sech}^2$
shape at a 40 MHz repetition rate and a temporal full width at half maximum around 0.7 ps at a tunable carrier wavelength of 1540 nm and 1560 nm. The main laser output is analyzed in the spectro-temporal domain using a free-space frequency-resolved optical gating device (Femto Easy Fast FROG PS10). An optical filter and a variable optical attenuator are placed after the source in order to control the spectral width of the pump and to ensure that we have a pump pulse as close as possible to a fundamental soliton. The pump field is then combined with a continuous wave probe with a wavelength adjusted to the group velocity matched wavelength of the pump field. The two pulses are injected into a $5$-km-long dispersion shifted fiber (DSF). 
The dispersion properties of the DSF were obtained via a time-of-flight measurement.
Fig.~\ref{fig:Db1} shows the measurement of the group velocity dispersion $\Delta\beta_1$ which we fit with a third order polynomial.
Its nonlinear coefficient is taken to be $\gamma = 1.8\ {\rm W}^{-1}.{\rm km}^{-1}$ and the dispersion coefficients are $\beta_2 = -1.52 \ {\rm ps}^{2}.{\rm km}^{-1}$, $\beta_3 = 0.12 \ {\rm ps}^{3}.{\rm km}^{-1}$ and $\beta_4 = -0.0006 \ {\rm ps}^{4}.{\rm km}^{-1}$ at the wavelength $\lambda = 1560 \ {\rm nm}$.
The output signal is characterised with an optical
spectrum analyzer.

\newpage
\baselineskip 11pt
\section*{\Large Supplementary information on the article: Analogue black hole lightring and its resonances in an optical fiber}

\section{Optical analogue of the black hole lightring}

\subsection{Eikonal approximation}

The evolution equation for a weak perturbation, $a(z,T)$, at the group velocity matched frequency, $\omega_{\rm GVM}$, satisfy the following linear PDE~\cite{Torres2026OL}:

\begin{equation}
    \frac{\partial a}{\partial z} - \sum_{m\geq 2} \frac{i^{(m+1)}}{m!} \beta^{\rm GVM}_m \frac{\partial^m a}{\partial T^m} = 2i\gamma |A_P(z,T)|^2 a(z,T).\label{eq:linear_WE}
\end{equation}
where $|A_P(z,T)|^2$ is the pump field intensity. The dispersion coefficient $\beta_m^{\rm GVM}$ are evaluated at the GVM frequency.
\\
The lightring frequency is obtain from the eikonal approximation to Eq.\eqref{eq:linear_WE}. We therefore look for solution with a rapidly oscillating phase and write $a(z,T) = R(z,T)\rm{exp}\left[ i S(z,T)/\epsilon \right]$. Assuming that the variations occurs on scales much larger than the wavelength, we rescale the derivative as $\partial \rightarrow \epsilon \partial$.
Performing a perturbative expansion, we get at leading order:
\begin{equation}
    \frac{\partial S}{\partial z} - \sum_{m\geq 2} \frac{(-1)^m}{m!}\beta^{\rm GVM}_m \left(\frac{\partial S}{\partial T}\right)^m = 2\gamma |A_P(z,T)|^2.\label{eq:eikonal}
\end{equation}
From the above equation, we define $k_z = \partial_z S $ and $\Omega = \partial_T S$ as well as the Hamiltonian:
\begin{equation}
    \mathcal{H}(z,T,k_z,\Omega) = k_z - \sum_{m\geq 2} \frac{(-1)^m}{m!}\beta^{\rm GVM}_m \Omega^m - 2\gamma |A_P(z,T)|^2.
\end{equation}
From the above Hamiltonian, we can obtain directly the trajectories of the \textit{rays} of the wave by solving Hamilton's equations
\begin{eqnarray}
\dot{T} = \frac{\partial \mathcal{H}}{\partial\Omega},\quad \dot{z} = \frac{\partial \mathcal{H}}{\partial k_z};\\
\dot{\Omega} = -\frac{\partial \mathcal{H}}{\partial T},\quad \dot{k_z} = -\frac{\partial \mathcal{H}}{\partial z}.
\end{eqnarray}
In the above, ``\ $\dot{}$\ ", refers to differentiation with respect to a parametrization of the rays. Since $\mathcal{H}$ is linear in $k_z$, we have that $\dot{z} = 1$ and therefore, $z$ can be used as an affine parameter for the rays. In the following, ``\ $\dot{}$\ " will refer to $\frac{d}{dz}$.
\subsection{Lightring}
Lightrings are defined from the above Hamiltonian in the specific case of $z$ independent pump fields. Under this condition, the lightring is defined as a critical point of the Hamiltonian for $(\Omega,T)$, i.e. it corresponds to an equilibrium point defined by
\begin{equation}
    \dot{\Omega} = -\frac{\partial\mathcal{H}}{\partial T} = 0\quad \text{and} \quad  \dot{T} = \frac{\partial\mathcal{H}}{\partial \Omega} = 0.
\end{equation}

Assuming that the temporal profile of the pump field is composed of a single pulse with a maximum centred at $T=0$, for example a fundamental soliton, the lightring corresponds to the point in the reduced phase space $(\Omega^{\rm LR},T^{\rm LR}) = (0,0)$.
Using the condition $\mathcal{H} = 0$, we can directly infer the lightring wavenumber 
\begin{equation}
    k_z^{\rm LR} = 2\gamma|A_P(0)|^2.\label{eq:LR_real}
\end{equation}
\subsection{Decay rate}
We have seen that the lightring is defined as an equilibrium point. In the case of black holes and the optical analogue of interest here, the lightring is an \textit{unstable} equilibrium, meaning that a small wavepacket located on the lightring will eventually leak out of the lightring. This exponential leak out is associated with the complex part of the lightring frequency that we now compute.
\\
We consider a family of rays close to the lightring $(k_z^{\rm LR} + \delta k_z,\Omega^{\rm LR} + \delta \Omega, T^{\rm LR} + \delta T) = (k_z^{\rm LR} + \delta k_z, \delta \Omega, \delta T)$. In the vicinity of the lightring, the Hamiltonian can be expanded as
\begin{equation}
    \mathcal{H}\approx \delta k_z - \frac{1}{2}\beta^{\rm GVM}_2 \delta\Omega^2 - \gamma I_P''(0) \delta T^2,
\end{equation}
where $I_P''(0) = d^2|A_P(T)|^2/dT^2$. We note that since $|A_P(T)|$ is maximum at the lightring $I_P''(0)<0$. In the case of a fundamental soliton where $|A_P(T)|^2 = P_S\ {\rm{sech}}^2(T/T_S)$, we have that $I_P''(0) = -2 P_S/T_S^2$.
From Hamilton's equation, we have
\begin{equation}
    \ddot{\delta T} = 2\gamma \beta^{\rm GVM}_2 |I''_P(0)| \delta T.
\end{equation}
The general solution is a linear superposition of decaying and growing exponentials. At large $z$, the growing behavior dominates, and we have
\begin{equation}
    \delta T \propto e^{\Lambda z},
\end{equation}
where $\Lambda = \sqrt{2\gamma \beta^{\rm GVM}_2 |I''_P(0)|}$ is the Lyapunov exponent governing the rate of exponential growth away from the orbit. It further governs the decay of waves close to the circular orbit. Indeed, the next-to-leading order of the eikonal approximation in the vicinity of the lightring gives the evolution of the amplitude $R$ along a ray
\begin{equation}
    \frac{d }{dz}\left( R^2\beta_2^{\rm GVM}\delta \Omega\right) = 0.
\end{equation}
Therefore, we have that 
\begin{equation}
    R \propto e^{-\frac{\Lambda}{2} z}. \label{eq:decay}
\end{equation}
Combining Eqs.~\eqref{eq:LR_real} and \eqref{eq:decay}, we define the lightring wavenumber as
\begin{equation}
    k_z^\star = k_z^{\rm LR} - i \frac{\Lambda}{2}.
\end{equation}
Experimentally, we do not have access to the evolution along $z$, therefore we translate the comples wavenumber into a complex frequency associated to the temporal profile which can be measured. This is done via the Hamiltonian away from the optical potential by solving
\begin{equation}
k_z^\star = \sum_{m\geq 2} \frac{(-1)^m}{m!}\beta^{\rm GVM}_m (\Omega^\star)^m\label{eq:LR_freq_gen}    
\end{equation}
In cases where one can neglect higher order dispersion, the above equation reduces to
\begin{equation}
\Omega^\star = \sqrt{\frac{2 k_z^\star}{\beta_2^{\rm GVM}}}.\label{eq:LR_freq}    
\end{equation}

\end{document}